\documentclass[12pt,preprintnumbers,nofootinbib,letterpaper]{article}
\pdfoutput=1
\usepackage{amsmath,amssymb,amsfonts,cite,xcolor,epsf,epsfig,epstopdf,graphics,booktabs,graphicx,mathtools,subcaption,verbatim}
\def\thefootnote{*\arabic{footnote}}
\definecolor{ultramarine}{rgb}{0.07, 0.04, 0.56}
\definecolor{cadmiumgreen}{rgb}{0.0, 0.42, 0.24}
\definecolor{indigo(dye)}{rgb}{0.0, 0.25, 0.42}
\usepackage[linktocpage=true,breaklinks]{hyperref}
\hypersetup{
	colorlinks=true,
	citecolor=ultramarine,
	linkcolor=cadmiumgreen,
	urlcolor=indigo(dye),
}

\usepackage[utf8]{inputenc}

\numberwithin{equation}{section}

\usepackage{array}
\newcolumntype{P}[1]{>{\centering\arraybackslash}p{#1}}

\usepackage[shortlabels]{enumitem}

\usepackage{dcolumn}
\newcolumntype{M}[1]{>{\centering\arraybackslash}m{#1}}
\newcolumntype{N}{@{}m{0pt}@{}}

\usepackage{amsmath}
\usepackage{ulem}
\usepackage{cancel}
\usepackage{makecell}

\newcommand{\Mpl}{M_{\rm Pl}}

\newcommand{\D}{{\rm d}}

\newcommand{\be}{\begin{equation}}  
	\newcommand{\ee}{\end{equation}}

\usepackage{tikz}

\begin{document}
	\noindent\hfill \text{CQUeST-2026-0770}
	
	\begin{center}
		
		\def\thefootnote{\fnsymbol{footnote}}
		
		\vspace*{1.5cm}
		{\Large {\bf A no-go theorem in bumblebee vector-tensor cosmology}}
		\\[1cm]
		
		{Carsten van de Bruck$^{1}$, Mohammad Ali Gorji$^{2}$, Nils A. Nilsson$^{2,3}$, Masroor C. Pookkillath$^{4} $, Masahide Yamaguchi$^{2,5,6}$}
		\\[.7cm]
		
		{\small \textit{$^{1}$School of Mathematical and Physical Sciences, University of Sheffield, Hounsfield Road, Sheffield S3 7RH, United Kingdom}}\\
		
		{\small \textit{$^2$Cosmology, Gravity, and Astroparticle Physics Group, Center for Theoretical Physics of the Universe, Institute for Basic Science (IBS), Daejeon, 34126, Korea
		}}\\
		
		{\small \textit{$^3$LTE, Observatoire de Paris, Université PSL, CNRS, LNE, Sorbonne Universit\'e, 61 avenue de l’Observatoire, 75 014 Paris, France}}\\
		
		{\small \textit{$^{4}$Center for Quantum Spacetime, Sogang University, 35 Baekbeom-ro, Mapo-gu, Seoul 04107, Korea}}\\
		
		{\small \textit{$^{5}$Department of Physics, Institute of Science Tokyo, 2-12-1 Ookayama, Meguro-ku, Tokyo 152-8551, Japan}}\\
		
		{\small \textit{$^{6}$Department of Physics \& Institute of Physics and Applied Physics (IPAP),Yonsei University, Seoul, 03722, Korea}}\\
		
	\end{center}
	
	\vspace{0.5cm}
	
	\hrule \vspace{0.5cm}
	
	\begin{abstract}
		Bumblebee models, a class of vector-tensor theories in which a vector field acquires a nonzero vacuum expectation value that spontaneously breaks spacetime symmetries, are ubiquitous in the literature. By constructing the most general bumblebee action from all diffeomorphism-invariant marginal operators together with a general potential, aiming to cover all the bumblebee models studied in the literature, we perform a complete linear perturbation analysis on a spatially flat FLRW background. We show that for generic marginal couplings, the scalar sector propagates extra degrees of freedom beyond the single scalar expected for a massive vector. Enforcing the correct number of propagating modes in a cosmological setup forces degeneracy relations between the marginal couplings, which in turn completely fix the potential at the background level and render the remaining scalar infinitely strongly coupled already at linear order of perturbations. We establish a no-go theorem stating that the following conditions cannot be simultaneously satisfied: (i) the most general marginal action, (ii) a homogeneous and isotropic background, (iii) no extra propagating degrees of freedom around a spatially flat FLRW background, and (iv) healthy cosmological perturbations.
	\end{abstract}
	
	\vspace{0.5cm} 
	
	\hrule
	\def\thefootnote{\arabic{footnote}}
	\setcounter{footnote}{0}

	\newpage

	\section{Introduction}\label{sec:intro}
	At the present time, we have achieved a detailed understanding of the Universe, and the techniques employed, be it theoretically or observationally, have reached high levels of sophistication. The open questions are a few, but their importance cannot be overemphasized. Puzzles such as the cosmological constant problem~\cite{Weinberg:1988cp,Martin:2012bt},  the nature of dark matter \cite{Marsh:2024ury} and dark energy~\cite{Huterer:2017buf}, and the Hubble tension~\cite{Schoneberg:2021qvd,DiValentino:2025sru} keep us from a truly elegant description of the Universe from primordial times until today. Recent cosmological surveys have also made it evident that measurements of quantities such as $H_0$ and $S_8$ give conflicting results (at $>5\sigma$ confidence level in the case of the Hubble tension) when measured with early-time probes as compared to late-time ones (see for example \cite{CosmoVerse:2025txj} and many references therein). Whether these issues are ultimately due to hitherto unknown systematics or to a fundamental misunderstanding of the action of gravity on cosmological scales, attempts to modify gravity are now strongly motivated.
	
	Modified gravity models containing vector fields, so-called vector-tensor theories, are natural extensions to general relativity beyond scalar fields. They also appear quite naturally in models with preferred spacetime foliations such as Einstein-Aether~\cite{Jacobson:2000xp,Eling:2004dk} (see also e.g.~\cite{Benisty:2021cin}) and other khronometric theories. In cosmology, a propagating vector mode at the background level can lead to anisotropy~\cite{Koivisto:2008ig,Koivisto:2008xf,Himmetoglu:2008zp,Watanabe:2009ct,Bartolo:2012sd,Ohashi:2013qba,Heisenberg:2016wtr,DeFelice:2025ykh,DeFelice:2025khe}, but models with multiple vectors~\cite{Golovnev:2008cf,Emami:2016ldl,Gorji:2018okn,Firouzjahi:2018wlp,Gorji:2020vnh} or non-Abelian extensions \cite{Maleknejad:2011jw,Maleknejad:2011sq,Galtsov:1991un,Gorji:2019ttx} allow for isotropic dynamical vector degrees of freedom. Moreover, even if the vector field is non-dynamical at the background level, it can still have non-trivial effects at the level of perturbations \cite{Mukohyama:2006mm,DeFelice:2016yws,DeFelice:2016uil,DeFelice:2020sdq,Aoki:2021wew,Aoki:2023bmz,Aoki:2024ktc,Tomizuka:2025dpy}. Many vector–tensor theories appear as subsets of generalized Proca theories~\cite{Heisenberg:2014rta}, which are extensions of the seminal Proca massive electrodynamics but do not change the number of propagating degrees of freedom (two transverse and one longitudinal) of the vector field $A^\mu$. In standard Proca theory, the temporal component $A^0$ of the vector field does not propagate, but instead appears as a primary constraint which is only first class when the vector field is massless; therefore, $A^0$ does not propagate in neither Proca nor generalized Proca.
	In generalized Proca, the longitudinal mode behaves as a scalar Galileon, which has implications for infrared modifications of gravity and falls under the Horndeski class, and in this sense generalized Proca can be regarded as a scalar-vector-tensor approach to modified gravity effective field theory (EFT) \cite{Mukohyama:2006mm,Aoki:2021wew,Aoki:2023bmz,Aoki:2024ktc,Tomizuka:2025dpy}. A significant amount of work has been done on aspects of generalized Proca, for example cosmological applications~\cite{DeFelice:2016yws,DeFelice:2016uil} including applications to cosmic tensions~\cite{CosmoVerse:2025txj,DeFelice:2020sdq, Keeley:2019esp}, positivity and causality~\cite{deRham:2018qqo}, compact objects, stellar structure~\cite{Olmo:2019flu,Babichev:2017rti}, and more.
	Since some of these models feature spacetime-symmetry breaking, we consider a generalised version of specific vector-tensor theories known as {\it bumblebee models} and which appear as vector subsets of the gravity sector of a commonly used EFTs known as the Standard-Model Extension (SME) and which is used to search for possible violations of Lorentz, diffeomorphisms, and {\it CPT} symmetry~(see \cite{Kostelecky:2003fs} for the gravity sector).\footnote{See \cite{Kostelecky:2008ts} for an exhaustive summary of current constraints in all sectors including gravity.} This EFT framework was first introduced by Kostelecky and Samuel in 1989~\cite{Kostelecky:1989jp,Kostelecky:1989jw}, only a year after their seminal work~\cite{Kostelecky:1988zi} showing the spontaneous breaking of local Lorentz symmetry in string theory, and its coupling to gravity was presented in for example~\cite{Kostelecky:2003fs}. The interest in bumblebee models has skyrocketed in recent years, particularly in the context of black holes where a large body of literature now exists,\footnote{A search at the time of writing reveals almost 200 papers written in the past ten years.} for example the now well-known Schwarzschild-like Casana solution~\cite{Casana:2017jkc}, other static solutions~\cite{Xu:2022frb,Filho:2022yrk,Bailey:2025oun}, rotating solutions~\cite{Ding:2019mal,Li:2020dln,Liu:2022dcn}, and many more; for example, see~\cite{Bailey:2013fwa,Lambiase:2023zeo,Bonder:2015jra,Ovgun:2018xys,Maluf:2020kgf,Jesus:2020lsv,Liang:2022hxd,Xu:2023xqh,PhysRevD.71.065008,Bluhm:2008yt}, which is by no means an exhaustive list. Work on cosmology includes an analysis of G\"odel solutions~\cite{Jesus:2020lsv} background Friedmann-Lemaître-Robertson-Walker (FLRW) and anti-de Sitter solutions~\cite{Reyes:2024hqi,Bailey:2025oun,Maluf:2020kgf}, tests with distance measures and cosmic microwave background temperature anisotropies~\cite{Xu:2025bvx,Zhu:2024qcm}. The bumblebee model has also attracted attention in the context of anisotropic cosmological solutions~\cite{Maluf:2021lwh} as well as for Kasner-type universes~\cite{Neves:2022qyb}.
	
	In this paper, we clarify several issues regarding the bumblebee model with non-minimal coupling to gravity; specifically, we study cosmological perturbations around a spatially flat FLRW background and show that, in general, the model contains three longitudinal modes. We then impose relations among the marginal constant couplings to remove the two extra unwanted degrees of freedom; these relations coincide with the degeneracy conditions, so that after imposing them the model reduces to a subset of generalized Proca theory. However, the remaining scalar mode becomes strongly coupled, leading to the no-go theorem that constitutes the main result of this paper.
	
	The paper is organized as follows: in Section~\ref{sec:generalized_bumb}, we introduce and set up the model under consideration, generalizing the standard bumblebee action by introducing all possible marginal operators, which includes the bumblebee models studied in the literature; 
	in Section~\ref{sec:bg}, we discuss the equations of motion assuming homogeneous and isotropic flat FLRW background;
	in Section~\ref{sec:cosmoperts}, we perform the cosmological perturbation analysis and discuss the stability issue of the model; in Section~\ref{sec:nogo}, we construct the no-go theorem; in Section~\ref{sec:disc}, we present our conclusions. Throughout the paper, we use the mostly-plus  metric signature ($-+++$) and we adopt the units $c=\hbar=1$ and $G=1/(8\pi M_{\rm Pl}^2)$, where $M_{\rm Pl}$ is the reduced Planck mass.
	
	\section{Systematic construction of bumblebee gravity}\label{sec:generalized_bumb}
	
	Bumblebee theories are most naturally interpreted as EFTs for a vector field $B_\mu$ whose vacuum expectation value is selected by a symmetry-breaking potential $V\!\left(B_\mu B^\mu \pm b^2\right)$, where $\langle B_\mu B^\mu \rangle =\mp b^2$. An important generic feature is that the presence of such a potential explicitly\footnote{Although the breaking of Lorentz and diffeomorphism symmetry is spontaneous.} breaks any local $U(1)$ gauge symmetry for $B_\mu$, independently of the detailed choice of kinetic term $B_{\mu\nu}B^{\mu\nu}$ with
	$B_{\mu \nu} \equiv \nabla_{\mu}B_{\nu} - \nabla_{\nu}B_{\mu}$. One is therefore not restricted to gauge-invariant interactions, and the vector-tensor realization of the bumblebee setup already includes non gauge-invariant but diffeomorphism-covariant operators such as $B_\mu B_\nu R^{\mu\nu}$ and $B_\mu B^\mu R$ together with the potential. Since $B_\mu$ has canonical mass dimension one in four spacetime dimensions, these curvature couplings sit at canonical dimension four and hence are marginal (in the EFT sense) at the two-derivative level, with dimensionless couplings, while the only intrinsic scale associated with spacetime-symmetry breaking enters through the potential via the vacuum condition $\langle B_\mu B^\mu \rangle =\mp b^2$. In this regard, we look for an extension of bumblebee gravity by including the complete basis of diffeomorphism-invariant marginal operators built from
	\begin{align}\label{eq:building-blocks}
		B_\mu \,, 
		\quad 
		g_{\mu\nu} \,, 
		\quad
		\nabla_\mu \,, 
		\quad 
		\epsilon^{\mu\nu \rho \sigma} \,,
		\quad
		R_{\mu\nu\rho\sigma} \,,
	\end{align}
	where $g_{\mu\nu}$ is the spacetime metric, $\nabla_\mu$ is the torsion-free, metric-compatible (Levi-Civita) covariant derivative,  $\epsilon^{\mu\nu\rho\sigma}$ is the totally antisymmetric Levi-Civita tensor, and $R_{\mu\nu\rho\sigma}$ is the Riemann tensor. The scale of spacetime-symmetry breaking, which controls the vacuum structure and symmetry-breaking pattern, enters only through the potential.
	
	With the above criteria as our guiding principle, we enumerate in Table~\ref{tab:buildingblocks} the complete set of diffeomorphism-invariant marginal operators that can be constructed from the building blocks \eqref{eq:building-blocks}.
	\begin{table}[h!]
		\centering
		\begin{tabular}{c c c}
			Operator & Mass dim. ($d$) & Derivative exp. ($D$) \\\midrule
			$B$ & 1 & 0\\
			$\nabla B$ & 2 & 1\\
			$R$ & 2 & 2 \\
			$B^2\nabla B$ & 4 & 1 \\
			$(\nabla B)^2$, $B^2R$ & 4 & 2 \\
			$R \nabla B$ & 4 & 3 
			\\
			$R^2$ & 4 & 4
			\\\bottomrule
		\end{tabular}
		\caption{Index-suppressed building blocks of the bumblebee model.}
		\label{tab:buildingblocks}
	\end{table}
	From this table, we can enumerate all possible terms as
	\begin{enumerate}
		\item Pure gravity sector ($d=4, D=4$): As it is well known, we have three independent operators
		\begin{equation}\label{eq:D4}
			R^2 \,,
			\qquad R_{\mu\nu}R^{\mu\nu} \,,
			\qquad
			E_4\equiv R_{\mu\nu\rho\sigma}R^{\mu\nu\rho\sigma}-4R_{\mu\nu}R^{\mu\nu}+R^2\,,
		\end{equation}
		where $E_4$ is the Gauss-Bonnet (Euler) density.

		\item Linear in the bumblebee field ($d=4, D=3$): We have two possible operators coupling to gravity
		\begin{equation}\label{eq:D3}
			\nabla_{\mu} B_{\nu} R^{\mu \nu} \,, \qquad \nabla_{\mu} B^{\mu} R \,.
		\end{equation}
		\item Quadratic in the bumblebee field ($d=4, D=2$): In this case, we can have both kinetic and interaction terms with the gravity sector, which are given by
		\begin{equation}\label{eq:D2}
			(\nabla_{\mu}B^{\mu})^{2} \,, \quad B_{\mu\nu} B^{\mu\nu} \,, \quad \nabla_{\mu} B_{\nu} \nabla^{\nu}B^{\mu} \,, \quad B_{\alpha}B^{\alpha} R \,, \quad R_{\mu \nu} B^{\mu} B^{\nu} \,, 
			\quad B_{\mu\nu} \tilde{B}^{\mu\nu} \,,
		\end{equation}
		where $\tilde{B}^{\mu\nu}=\tfrac{1}{2}\epsilon^{\mu\nu\alpha\beta}B_{\alpha\beta}$ is the dual of $B_{\mu\nu}$. Note that $\nabla_{\mu}B_{\nu}\nabla^{\mu}B^{\nu}=B_{\mu \nu} B^{\mu\nu}/2+\nabla_{\mu}B_{\nu}\nabla^{\nu}B^{\mu}$ is not an independent building block.
		
		\item Cubic in the bumblebee field ($d=4, D=1$): In this case, there is no direct coupling with the curvature tensor since its mass dimension is two. Then, the possible terms are
		\begin{equation}\label{eq:cubicTerms}
			B_{\nu}B^{\nu} \nabla_{\mu} B^{\mu} \,, \qquad B_{\mu} B_{\nu} \nabla^{\mu} B^{\nu}\,. 
		\end{equation}
	\end{enumerate}
	
	All of the interactions listed above should therefore be included in the action, each multiplied by an independent {\it constant}, {\it dimensionless} coefficient. However, they are not all independent: we are free to perform integration by parts to identify redundancies and determine which operators are equivalent up to total derivatives. Let us consider the cases $D=1,2,3$ in turn.
	
	For ($d=4, D=4$), Lovelock's theorem implies that the only curvature-squared combination yielding second-order equations of motion is the Gauss-Bonnet density $E_4$ \cite{Lovelock:1971yv,Lovelock:1972vz}, and hence it does not introduce additional propagating modes associated with higher-derivative dynamics. However, in four dimensions \(E_4\) is topological (a total derivative). With a constant coefficient, it does not contribute to the equations of motion, so we omit it. Consequently, there are no relevant \(d=4\), \(D=4\) operators in the pure gravitational sector Eq.~\eqref{eq:D4} for our purposes.
	
	For ($d=4, D=3$) operators, performing integration by parts and using the Bianchi identity, we find
	\begin{align}\label{eq:identity2}
		\nabla_\mu B_\nu R^{\mu\nu}=\frac{1}{2}\nabla_\mu B^\mu R \,,
	\end{align}
	which implies that only one of the two operators in Eq.~\eqref{eq:D3} is independent, and we choose to drop $\nabla_\mu B_\nu R^{\mu\nu}$.
	
	For ($d=4, D=2$) operators, we have the following identity up to a boundary term
	\begin{align}\label{eq:identity1}
		& R_{\mu \nu} B^{\mu} B^{\nu} = \big( \nabla_{\mu}B^{\mu} \big)^2 - \nabla_{\mu} B_{\nu} \nabla^{\nu} B^{\mu} \,,
	\end{align}
	which implies that only five of the six operators in Eq.~\eqref{eq:D2} are independent, and we therefore drop $\nabla_{\mu} B_{\nu}\nabla^{\nu} B^{\mu}$. Moreover, since we deal with marginal operator with constant coefficeints, the last operator in Eq.~\eqref{eq:D2} is a total derivative. We thus do not need to include it in the action. Hence, there is no parity-violating operator. 
	
	For ($d=4, D=1$) operators, performing integration by parts we find
	\begin{equation}
		B_{\mu} B_{\nu} \nabla^{\mu} B^{\nu} = -\frac{1}{2} B_{\nu}B^{\nu} \nabla_{\mu} B^{\mu} \,,
	\end{equation}
	which shows that only one of the two operators in Eq.~\eqref{eq:cubicTerms} is independent, and we choose to drop $B_{\mu} B_{\nu} \nabla^{\mu} B^{\nu}$.
	
	After removing the redundant terms, we can write a general action with marginal operators of a vector-tensor theory (on a torsion-free Riemannian spacetime), with the pure gravitational sector given by the Einstein-Hilbert term, of the form
	\begin{equation}\label{eq:lagr}
		\begin{split}
			S_B = \int \D^4x \sqrt{-g} \Bigg[ &\frac{M_{\rm Pl}^{2}}{2} R - \frac{\gamma}{4}B_{\mu \nu}B^{\mu \nu} - V(B^2) + \eta B^2 \nabla_{\mu}B^{\mu} 
			\\
			&+ \xi B^\mu B^\nu R_{\mu\nu} +\sigma B^2 R 
			+ \varsigma (\nabla_{\mu}B^{\mu} )^{2} 
			+ \upsilon 
			\nabla_\mu B^\mu R \Bigg] \,,
		\end{split}
	\end{equation}
	where all coupling coefficients $\{ \gamma, \eta, \xi, \sigma, \varsigma, \upsilon\}$ are dimensionless constants and $B^{2} \equiv B_{\mu}B^{\mu}$. We can set $\gamma=1$ without loss of generality; however, we keep it explicit to keep track of this coupling.
	
	Various bumblebee models can be obtained as subsets of the general action \eqref{eq:lagr} by imposing relations between the coefficients $\{ \gamma, \eta, \xi, \sigma, \varsigma, \upsilon\}$; in particular, by imposing $\{\varsigma\to0\,,\eta\to0,\,\gamma\to1, \xi\to M_{\rm Pl}^2\xi/2\}$, we obtain a popular bumblebee model with non-minimal couplings $\xi B^\mu B^\nu R_{\mu\nu} +\sigma B^2 R$
	which has been the subject of many studies (see for example \cite{Ovgun:2018xys,Casana:2017jkc,Li:2020dln,Maluf:2020kgf,Bertolami:2005bh,Ovgun:2018ran,Ding:2019mal,Kanzi:2019gtu,Xu:2025bvx,Li:2025bzo,Khodadi:2025wuw,Liang:2022hxd,Khodadi:2021owg,Gullu:2020qzu,Ding:2020kfr,Reyes:2024hqi,Kostelecky:2003fs}).
	We also note that there are other subsets of our action \eqref{eq:lagr} in the literature, all of which can be related to the results above using simple parameter maps, for example~\cite{Gonzalez-Espinoza:2025fmi,Lai:2025nyo}; we display a selection of such maps in Table~\ref{tab:maps}.
	\begin{table}[h]
		\def\arraystretch{1.2}
		\centering
			\begin{tabular}{ccc}
				\toprule
				Operators & Reference & Map\\
				\midrule
				$BBR$ & e.g. \cite{Kostelecky:2003fs} & \makecell{$\xi\to M_{\rm Pl}^2\,\xi/2\,, \gamma\to 1\,, \sigma\to 0\,, \varsigma\to0\,,\eta\to 0$.}\\[2mm]
				\makecell{$BBR$, $\nabla B\nabla B$} & \cite{ Bluhm:2007bd, Gonzalez-Espinoza:2025fmi} & $\eta \to0 \,, \gamma \to \tau_{1} - \tau_{2}\,, $
				$\xi = \sigma_{1} -(\tau_{2} + \tau_{3})/2\,$,\\&& $\sigma \to \sigma_{2}\,, \varsigma \to (\tau_{2} + \tau_{3})/2\,.$ \\[2mm]
				$BBR$, $\nabla B\nabla B$ & \cite{Lai:2025nyo} & \makecell{$\varsigma\to0,\,\gamma\to1 ,\,\sigma\to 0,\,$}
				$ \eta \to 0$, $\xi \to \Mpl^2
				\,\xi/2 \,.$ \\
				\bottomrule
			\end{tabular}
		\caption{Bumblebee subsets and their maps to our general construct.}
		\label{tab:maps}
	\end{table}
	
	From the action \eqref{eq:lagr} we obtain the following covariant equations of motion for the metric
	\begin{equation}\label{eq:generalized_einstein}
		M_{\rm P}^2 \, G_{\mu\nu}+ \gamma \left(\frac{1}{4}g_{\mu\nu}B_{\alpha\beta}B^{\alpha\beta}-B_{\mu}^{~\alpha}B_{\nu\alpha} \right) + g_{\mu\nu}V(B^2) - 2B_{\mu}B_{\nu} V^{\prime} +\sum_i \alpha_i (\mathcal{X}_i)_{\mu\nu} = 0,
	\end{equation}
	where $G_{\mu\nu} = R_{\mu\nu}- g_{\mu\nu}R/2$
	is the Einstein tensor and $\alpha_i \in \{\xi, \sigma, \varsigma,\eta, \upsilon\}$. The explicit forms of $(\mathcal{X}_i)_{\mu\nu}$ are displayed in Appendix~\ref{app:chis}. We also find the generalized bumblebee equation of motion as,
	\begin{equation}\label{eq:Bmu}
		\begin{aligned}
			\frac{\gamma}{2}\Box B_\mu-\frac{1}{2}(\gamma+2\varsigma)\nabla_\nu\nabla_\mu B^\nu+(\xi+\varsigma)B^\nu R_{\mu\nu}-\eta B^\nu\nabla_\mu B_\nu+\sigma B_\mu R+\eta B_\mu\nabla_\nu B^\nu \\
			-\frac{\upsilon}{2} \nabla_{\mu} R -B_\mu V^\prime(B^2) = &0 \,,
		\end{aligned}
	\end{equation}
	where $\Box \equiv \nabla_\lambda\nabla^\lambda$ and where
	a prime denotes derivative with respect to the argument. 
	
	The bumblebee model is characterized by the bumblebee field $B_\mu$ acquiring a non-zero background value, commonly called the vacuum expectation value (vev), denoted by
	\begin{align}
		\bar{B}_\mu=\langle B_\mu \rangle \,.
	\end{align}
	In flat space, the vev is constant such that $\bar{B}_\mu=\mbox{constant}$ is the solution of the system. This can be achieved by an appropriate choice of the potential $V(B^2)$ which triggers $\bar{B}_\mu =\mbox{constant}$. This is because $R_{\mu\nu}=0$ in flat space and Eq.~\eqref{eq:Bmu} simply implies $V^\prime=0$. We can therefore write the potential as $V = V(B^2 \pm b^2)$ where $b$ is a constant which characterizes the amplitude of $\bar{B}_\mu$.\footnote{The potential should be at least quadratic in $B^2$ such that $V'=0$ implies $B^2=\mp{b}^2$. Depending on the choice of background for $B_\mu$, it can be timelike or spacelike, and we may have to choose either the plus or minus sign.}
	In curved spacetime, $\bar{B}_\mu$ can be a general four-vector $\bar{B}_\mu(x)$ which depends on the spacetime coordinates. The non-zero background $\bar{B}_\mu(x)$ then triggers the spontaneous breaking of diffeomorphism invariance. The metric can be brought to the Minkowski form at each point on the spacetime manifold, $g_{\mu\nu}(x)=e_\mu{}^{\bar{a}}(x)e_\nu{}^{\bar{b}}(x)\eta_{\bar{a}\bar{b}}$ where $\bar{a},\bar{b}$ are local Lorentz indices and $\eta_{\bar{a}\bar{b}}=\mbox{diag}(-1,1,1,1)$ with bars over components indicating the locally Lorentz frame.
	Let us first assume that $\bar{B}_\mu$ has a constant magnitude even in curved spacetime, and consider a timelike vector $b_{\bar{a}}=(b,\vec{0})$ in local Lorentz space with constant $b$: at the point ${\cal P}$ on the spacetime manifold, we can always use the freedom in the local Lorentz transformations to choose the vierbein such that $\bar{B}_\mu(x)=e_\mu{}^{\bar{a}}(x) b_{\bar{a}}=b\,e_\mu{}^{\bar{{0}}}(x)$. It is then follows that $\bar{B}^\mu\bar{B}_\mu = e_\mu{}^{\bar{a}} e^{\mu \bar{c}} b_{\bar{a}} b_{\bar{c}} = \eta^{\bar{a}\bar{c}} b_{\bar{a}}b_{\bar{c}}=b^2$.
	Although it is, in principle, possible, we do not need to assume that $\bar{B}_\mu(x)$ has a constant magnitude in curved spacetime; indeed, this assumption turns out to be too restrictive. In our setup, Eq. \eqref{eq:Bmu} shows that the non-minimal couplings can induce an effective mass for the bumblebee field, and  consequently the effective potential of the bumblebee field depends on the spacetime curvature $R$, which is not generally constant. Therefore, the magnitude of $\bar{B}_\mu(x)$ should in general depend on the spacetime coordinates. In the next section, we focus on the FLRW background, where the spacetime curvature depends on time, and therefore the bumblebee vev $\bar{B}_\mu$ must also depend on time.
	
	\section{Cosmological background evolution}\label{sec:bg}
	We consider the spatially flat FLRW metric in cosmic time $t$, which takes the form
	\begin{equation}\label{metric-FLRW}
		\D{s}^2 = -\D{t}^2 + a^2(t)\delta_{ij}\D{x}^i \D{x}^j,
	\end{equation}
	where $a$ is the scale factor. We adopt the corresponding vierbein components
	\begin{equation}\label{FLRW-tetrads}
		e_t{}^{\bar{0}}=1, \quad e_t{}^{\bar{i}}=0, \quad e_i{}^{\bar{0}}=0, \quad e_i{}^{\bar{j}}=\delta_i{}^{\bar{j}}a \,,
	\end{equation} 
	where we have shown the spatial local Lorentz indices as $\bar{i}, \bar{j}$ to distinguish them from the spacetime spatial indices $i,j$. To keep our ansatz as general as possible, $\bar{B}_\mu$ should depend on spacetime coordinates, and since the FLRW background \eqref{metric-FLRW} is homogeneous, $\bar{B}_\mu$ can depend only on time and have only a temporal component due to isotropy, leaving us with the background configuration
	\begin{equation}\label{eq:bbpert}
		\bar{B}_{\mu} = (\bar{B}_0(t), \vec{0}) \,.
	\end{equation}
	For the metric \eqref{metric-FLRW} and background configuration \eqref{eq:bbpert}, we find $\bar{B}_\mu \bar{B}^\mu=-\bar{B}_0(t)^2$.
	In the local Lorentz frame, $b_{\bar{a}}= e_{\bar{a}}{}^\mu\bar{B}_{\mu}=e_{\bar{a}}{}^0\bar{B}_{0}(t)=\delta_{\bar{a}}^0\bar{B}_{0}(t)$ giving $b_{\bar{a}} b^{\bar{a}} = -\bar{B}_0(t)^2$. Therefore, even in the local Lorentz frame, the magnitude of $b_{\bar{a}}$ is time-dependent $b_{\bar{a}}(t)$.
	Since $\bar{B}_\mu\bar{B}^\mu$ is not constant, its evolution should be determined through the dynamics of the system. Considering this background configuration, we obtain the Friedmann equations from the Einstein equations \eqref{eq:generalized_einstein} as
	\begin{align}
		\nonumber 3M_{\rm Pl}^2H^2-6\eta \bar{B}_0^3H+9(2\sigma-\varsigma)\bar{B}_0^2H^2-3(\xi+2\sigma)H(\bar{B}_0^2)^{\dot{}}-\varsigma\dot{\bar{B}}_0^2+6(\xi+2\sigma+\varsigma&)\bar{B}_0^2\dot{H}\\ +6 \upsilon \Big[ \dot{\bar{B}}_{0}  \dot{H} -H \ddot{\bar{B}}_{0} - 2 H ^2 \dot{\bar{B}}_{0} +\bar{B}_{0}  \big(3  H ^3 -\ddot{H} -4 H  \dot{H} \big)\Big] \nonumber \\+2\varsigma \bar{B}_0\ddot{\bar{B}}_0-V(-\bar{B}_0^2)-2\bar{B}_0^2V^\prime(-\bar{B}_0^2)&=0\,,\label{eq:FE1}\\
		\nonumber M_{\rm Pl}^2(3H^2+2\dot{H})-2(\xi+2\sigma+\varsigma)\bar{B}_0\ddot{\bar{B}}_0-(2\xi+4\sigma+\varsigma)\dot{\bar{B}}_0^2-(2\xi+2\sigma+3\varsigma)(2\bar{B}_0^2\dot{H})\\
		+2 \upsilon \Big[-\tfrac{\D ^{3}}{\D t^{3}} \bar{B}_{0} -5 H  \ddot{\bar{B}}_{0}-\dot{\bar{B}}_{0} 
		\big(5  \dot{H} +3 H ^2 \big)+9 \bar{B}_{0}  H 
		\big( \dot{H} +H ^2\big)\Big] \nonumber \\-(2\xi+2\sigma+3\varsigma)(2H(\bar{B}_0^2)^{\dot{}}+ 3 \bar{B}_0^2H^2) - 2 \eta  \bar{B}_{0}^{2} \dot{\bar{B}}_{0}-V(-\bar{B}_0^2) & =0
		\label{eq:FE2}
	\end{align}
	and similarly the equation of motion for the bumblebee field Eq.~\eqref{eq:Bmu} gives
	\begin{equation}\label{eq:VFE}
		\varsigma(\ddot{\bar{B}}_0+3H\dot{\bar{B}}_0) + 3(\xi+2\sigma+\varsigma)\bar{B}_0\dot{H}+3(\xi+4\sigma)\bar{B}_0H^2-3\eta \bar{B}_0^2H-\bar{B}_0V^\prime(-\bar{B}_0^2) -3 \upsilon \big(\ddot{H} +4 H  \dot{H} \big)=0 \,.
	\end{equation}
	We note that the temporal component $\bar{B}_0$ of the bumblebee field propagates at the background level. The couplings $\{\xi,\sigma,\varsigma\}$ generate terms containing second time derivatives of the background field, $\ddot{\bar{B}}_{0}$, whereas the coupling $\upsilon$ generates terms with third time derivatives, $\dddot{\bar{B}}_{0}$. This is consistent with the operator classification in Table~\ref{tab:maps}: the former arise from $D=2$ operators while the latter originates from a $D=3$ operator; by contrast, the coupling $\eta$ does not render the temporal component dynamical. Consequently, in the absence of the derivative couplings, $\xi=\sigma=\varsigma=\upsilon=0$, the temporal component of the bumblebee field does not propagate at the background level. We also note that for the limit $\varsigma=\upsilon=0$, Eq.~\eqref{eq:VFE} turns into a constraint equation.

	\section{Linear cosmological perturbations}\label{sec:cosmoperts}
	To determine if the theory is stable around cosmological backgrounds, we extend our analysis to the linear perturbation regime, and we decompose the perturbations into scalar (S), vector (V), and tensor (T) modes. The action is invariant under the change of the coordinates (diffeomorphism invariance) $x^\mu\to{x}^\mu+\zeta^\mu$. The decomposition based on the symmetries of the FLRW metric is $\zeta^\mu=(\zeta^0,\partial^i\zeta+\zeta^{(T)\,i})$, where $\zeta^0$ and $\zeta$ transform as scalars under time-dependent spatial diffeomorphisms while the two transverse degrees of freedom $\zeta^{(T)\,i}$, which are divergenceless $\partial_i\zeta^{(T)\,i}=0$, transform as vectors. We choose these four gauge degrees of freedom such that the perturbed metric takes the form \cite{Mukhanov:2005sc,Weinberg:2008zzc}
	\begin{align}\label{metric-dec}
		\D s^2 = - \left( 1+2\alpha \right) \D t^2 + 2 a \left(\partial_i\beta+\beta_i\right) \D t \D x^i+ a^2\left(\delta_{ij}+ h_{ij}\right) \D x^i \D x^j \,.
	\end{align}
	We have two scalar modes ($\alpha, \beta$), two divergenceless vector modes ($\beta^i$), and two tensor modes ($h_{ij}$) which are symmetric, transverse $\partial^ih_{ij}=0$, and traceless $\delta^{ij}h_{ij}=0$.
	For the scalar perturbations, the spatial part of the metric is flat, earning this gauge the name spatially flat gauge. For the bumblebee field we have the general form
	\begin{align}\label{B-dec}
		B_{\mu} = (\bar{B}_0+\delta B_0, \,\, \partial_i\delta B_s+\delta B_i^{\perp}) \,,
	\end{align}
	where we note the existence of two scalar modes $(\delta B_0, \delta B_s)$ and two divergenceless vector modes $(\delta B_i^{\perp})$ at the level of perturbation. 
	
	We now substitute Eqs.~\eqref{metric-dec} and \eqref{B-dec} into the action (\ref{eq:lagr}) and expand it up to second order in perturbations (linear regime). As is well known, the scalar, vector, and tensor modes decouple at the linear level, and we therefore study them separately below.
	
	\subsection{Tensor perturbations}
	The quadratic action for the tensor modes reads
	\begin{equation}
		S_{B,T}^{(2)}=\frac{M_{\rm Pl}^2}{8}\int \D^3x \, \D{t} \, a^{3}\mathcal{K}_T\left(\dot{h}_{ij}\dot{h}^{ij}-\frac{c_T^2}{a^2}\partial_i h_{jk}\partial^i h^{jk}\right),
	\end{equation}
	where we have defined the kinetic coefficient function and sound speed as
	\begin{equation}\label{eq:tensorNoGhost}
		\mathcal{K}_T\equiv  \Big[1-2(\xi+\sigma)\tilde{B}_0^2 - \frac{2 \upsilon}{\Mpl} (\dot{\tilde{B}}_0+3H \tilde{B}_0) \Big], \quad c_T^2 \equiv \frac{1}{\mathcal{K}_T} \Big[1-2\sigma\tilde{B}_0^2- \frac{2 \upsilon}{\Mpl} (\dot{\tilde{B}}_0+3H \tilde{B}_0) \Big] \,,
	\end{equation}
	and where the dimensionless variable $\tilde{B}_0$ is defined as
	\begin{equation}
		\tilde{B}_{0} \equiv \frac{\bar{B}_{0}}{\Mpl} \,.
	\end{equation}
	Once $\xi\neq0$, $c_T$ is controlled by a combination of $\xi$, $\sigma$, and $\upsilon$. From gravitational wave observations, we have $|c_T-1|<{\cal O}(10^{-15})$ at late time \cite{LIGOScientific:2017zic}, which can be translated onto a constraint on $\xi$, $\sigma$, and $\upsilon$. However, for $\xi=0$, $c_T^2=1$ and becomes independent of $\sigma$ and $\upsilon$. Therefore, for small coupling limit, $c_T^2$ becomes
	\begin{equation}
		c_T^2 \approx 1 + 2\xi\tilde{B}_0^2 \,;
		\qquad
		\xi,\sigma,\upsilon \ll 1 \,,
	\end{equation}
	which is completely controlled by $\xi$. In that case, gravitational wave observations only put constraint on $\xi$.
	
	In order to avoid ghost and gradient instabilities we must also demand that
	\begin{equation}\label{eq:stability-tensor}
		\mathcal{K}_T>0, \qquad c_T^2>0 \,.
	\end{equation}
	Note that depending on the dynamics of the system and values of the coupling, the speed of tensor modes can be subluminal or superluminal.

	\subsection{Vector perturbations}
	For the vector sector, the gravitational vector modes ${\beta}^i$ are non-dynamical fields and can be integrated out through their equations of motion. After some integration by parts, the action reduces to
	\begin{equation}
		\begin{aligned}
			S^{(2)}_{B,V}
			=&\frac{\gamma}{2}\int \D^3x \, \D{t} \, a \bigg( \delta\dot{B}^\perp_i\,\delta\dot{B}^{\perp{i}}-\frac{c_V^2}{a^2}\partial_i\delta B^\perp_j \partial^i\delta B^{\perp{j}} \bigg),
		\end{aligned}
	\end{equation}
	where the sound speed takes a complicated form in general, and we omit it here. In the high-momentum limit, it reads
	\begin{equation}
		c_V^2 \approx 1+\frac{2\xi^2\tilde{B}_0^2}{\gamma{\cal K}_T} \,;
		\qquad
		k \gg aH \,,
	\end{equation}
	where $\mathcal{K}_{T}$ is the kinetic coefficient for the tensor modes. In order to ensure stability, we impose
	\begin{equation}
		\gamma>0, \qquad c_V^2>0 \,.
	\end{equation}
	Note that once we consider $\gamma>0$, the stability of the tensor modes Eq.~\eqref{eq:stability-tensor}, automatically implies $c_V^2>0$, showing that the vector modes are always superluminal in the high-momentum limit.
	
	\subsection{Scalar perturbations}
	For the scalar perturbations we work in Fourier space. After performing some integration by parts, it is straightforward to show that $\beta$ is non-dynamical and the quadratic action includes $\ddot{\alpha}$ and $\ddot{\delta{B}}_0$ which cannot be removed by further integration by parts. All these second time derivative terms originate from the coupling $\upsilon$ which is third-order in derivative expansion ($D=3$ operator in Table~\ref{tab:buildingblocks}). It is then useful to separate the contributions proportional to  $\upsilon$ from the rest of the couplings. Integrating out $\beta$, the reduced quadratic action can be written as
	\begin{equation}\label{eq:s2scalar-0}
		S_{B,S}^{(2)}= \int\frac{\D^3k}{(2\pi)^3}\D{t} \left(
		{ \mathcal{L}}^{(2)}_2 + { \mathcal{L}}^{(2)}_3 
		\right) \,,
	\end{equation}
	where the Lagrangian ${ \mathcal{L}}^{(2)}_3$ includes contribution from the coupling $\upsilon$ such that
	\begin{align}
		{ \mathcal{L}}^{(2)}_3 \big{|}_{\upsilon=0}= 0 \,.
	\end{align} 
	We do not present the explicit forms of ${ \mathcal{L}}^{(2)}_{2,3}$ here as they are involved. Considering the following field redefinition
	\begin{align}
		\tilde{\alpha} \equiv \alpha - \frac{\delta{B}_0}{\bar{B}_0} \,,
	\end{align}
	and performing some appropriate integration by parts, we can get rid of $\delta \ddot{B}_0$ after which we end up with contributions of the form
	\begin{align}
		{ \mathcal{L}}^{(2)}_2 &= { \mathcal{L}}^{(2)}_2\big(\tilde{\alpha},\dot{\tilde{\alpha}},\delta{B}_0, \dot{\delta{B}}_0, \delta{B}_s, \dot{\delta{B}}_s\big) \,,
		\\
		{ \mathcal{L}}^{(2)}_3 &= { \mathcal{L}}^{(2)}_3\big(\tilde{\alpha},\dot{\tilde{\alpha}},\ddot{\tilde{\alpha}},\delta{B}_0, \dot{\delta{B}}_0, \delta{B}_s, \dot{\delta{B}}_s\big) \,.
	\end{align}
	In order to deal with $\ddot{\tilde{\alpha}}$, we define a new variable
	\begin{align}\label{eq:Q-def}
		Q \equiv \dot{\tilde{\alpha}} 
		\qquad
		\Rightarrow
		\qquad
		\dot{Q} = \ddot{\tilde{\alpha}} \,.
	\end{align}
	We then write an equivalent Lagrangian
	\begin{align}\nonumber
		\mathcal{L}^{(2)}_{\rm tot} &\equiv \mathcal{L}^{(2)}_2 + \mathcal{L}^{(2)}_3 + \lambda \left(
		Q - \dot{\tilde{\alpha}}
		\right) 
		\\
		&= \mathcal{L}^{(2)}_2 + \mathcal{L}^{(2)}_3 + \lambda
		Q + \dot{\lambda} \tilde{\alpha} + {\rm T.D.}
		\,,
	\end{align}
	where in the second equality we have performed  integration by parts and ${\rm T.D.}$ denotes a total derivative term. In the above, equation of motion of $\lambda$ guarantees that Eq.~\eqref{eq:Q-def} holds. The original field $\tilde{\alpha}$ now becomes non-dynamical and we can integrate it out to find
	\begin{align}\label{eq:L-normal}
		{ \mathcal{L}}^{(2)}_{\rm tot} = { \mathcal{L}}^{(2)}_{\rm tot}\big(\dot{\lambda}, \lambda, \dot{Q}, Q, \delta{B}_0, \dot{\delta{B}}_0, \delta{B}_s, \dot{\delta{B}}_s\big) \,.
	\end{align}
	As can be seen, instead of working with $\{\ddot{\tilde{\alpha}},\dot{\tilde{\alpha}},{\tilde{\alpha}}\}$, we can work with  $\{\dot{\lambda},\lambda,\dot{Q},Q\}$ which have only first time derivatives. 
	
	Having transformed the reduced Lagrangian into the normal form \eqref{eq:L-normal} we can look at the kinetic terms to see how many degrees of freedom are propagating in our model. Rewriting the Lagrangian in matrix form we have
	\begin{equation}\label{eq:s2scalar-4t4}
		\mathcal{L}^{(2)}_{\rm tot} = \frac{1}{2}a^3\left(
		\dot{\mathcal{V}}_4^\dagger\mathbf{K}_4\dot{\mathcal{V}}_4 
		+\dot{\mathcal{V}}_4^\dagger\mathbf{N}_4\mathcal{V}_4 
		-\mathcal{V}_4^\dagger\mathbf{X}_4\mathcal{V}_4
		\right),
	\end{equation}
	where $\mathbf{K}_4$, $\mathbf{N}_4$, and $\mathbf{X}_4$ are the kinetic, friction, and generalized gradient $4\times4$ matrices respectively and where
	\begin{equation}
		\mathcal{V}_4^\dagger \equiv (\lambda,Q,\delta B_0,\delta B_s) \,,
	\end{equation}
	from which we find that we have
	\begin{align}
		\operatorname{max[rank}(\mathbf{K}_4)] = 3 \,.
	\end{align}
	This shows that there is a non-trivial cancellation taking place in the kinetic matrix $\mathbf{K}_4$ since this is $4\times4$ and the rank could be $4$ in general. We can show that the rank reduces in the following way
	\begin{align}
		\operatorname{max[rank}(\mathbf{K}_4)|_{\upsilon=0}] = 2 \,,
		\qquad
		\operatorname{max[rank}(\mathbf{K}_4)|_{\eta=\xi=\sigma=\varsigma=0}] = 3 \,,
	\end{align}
	which shows the coupling $\upsilon$ is responsible for the highest rank. This coupling is third-order in derivative expansion ($D=3$ in Table~\ref{tab:buildingblocks}) and it provides three propagating degrees of freedom; however, we know that a healthy theory with a massive vector field should have one propagating scalar degree of freedom. We thus need to impose that $\upsilon$ vanishes
	\begin{align}\label{eq:upsilon-zero}
		\boxed{\upsilon = 0} \,.
	\end{align}
	Imposing now Eq.~\eqref{eq:upsilon-zero}, $\mathcal{L}^{(2)}_{\rm tot}$ reduces to $\mathcal{L}^{(2)}_2$ which in terms of $\alpha$ can be written as
	\begin{equation}\label{eq:s2scalar}
		\mathcal{L}^{(2)}_2 = \frac{1}{2}a^3\left(
		\dot{\mathcal{V}}_3^\dagger\mathbf{K}_3\dot{\mathcal{V}}_3 
		+\dot{\mathcal{V}}_3^\dagger\mathbf{N}_3\mathcal{V}_3 
		-\mathcal{V}_3^\dagger\mathbf{X}_3\mathcal{V}_3
		\right),
	\end{equation}
	with the variables 
	\begin{equation}
		\mathcal{V}_3^\dagger \equiv (\alpha,\delta B_0,\delta B_s) \,.
	\end{equation}
	In this case, the kinetic matrix takes the simple form
	\begin{equation}\label{eq:Kmat1}
		\mathbf{K}_3=
		\begin{pmatrix}
			K \bar{B}_0^2 & -K \bar{B}_0 & 0 \\
			-K \bar{B}_0 & K & 0 \\
			0 & 0 & \gamma\frac{k^2}{a^2}
		\end{pmatrix}\,,
	\end{equation}
	where we have defined the quantity $K$ as
	\begin{equation}\label{eq:def-K}
		K \equiv -\frac{2}{\varsigma} (\xi + 2 \sigma )(\xi +2\sigma+2\varsigma) \,.
	\end{equation}
	From Eq.~\eqref{eq:Kmat1} we find
	\begin{align}
		\operatorname{rank}(\mathbf{K}_3) = 2 \,,
	\end{align}
	which shows that there are two propagating degrees of freedom in general. The coupling $\eta$ does not play any role in the kinetic matrix and is therefore totally safe to include since it does not lead to any extra propagating degree of freedom. 
	
	Looking at Eq.~\eqref{eq:Kmat1} and \eqref{eq:def-K}, we see that the situation is different from the case of the third-order derivative operator $\upsilon$, where we only had one trivial choice of \eqref{eq:upsilon-zero} for a healthy theory; here, there are nontrivial possibilities which reduces the rank of $\mathbf{K}_3$ from 2 to 1. 
	From Eq.~\eqref{eq:def-K}, we see that $\mathbf{K}_3$ is singular for $\varsigma=0$; however, from the action \eqref{eq:lagr} we know that $\varsigma=0$ should not be a singular limit. This apparent singularity shows up since $\varsigma$ appears in the denominator of the solution for the non-dynamical field $\beta$. We thus need to first set $\varsigma=0$ in the quadratic Lagrangian and then integrate out $\beta$. Doing so, we will find that we can indeed integrate not only $\beta$ but also $\alpha$ and we find the following result
	\begin{equation}\label{eq:s2scalar}
		\mathcal{L}^{(2)}_2\big{|}_{\varsigma=0} = \frac{1}{2}a^3\left(
		\dot{\mathcal{V}}_2^\dagger\mathbf{K}_2\dot{\mathcal{V}}_2 
		+\dot{\mathcal{V}}_2^\dagger\mathbf{N}_2\mathcal{V}_2
		-\mathcal{V}_2^\dagger\mathbf{X}_2\mathcal{V}_2
		\right),
	\end{equation}
	with the variables 
	\begin{equation}
		\mathcal{V}_2^\dagger \equiv (\delta B_0,\delta B_s) \,,
	\end{equation}
	and
	\begin{equation}\label{eq:Kmat1-varsigma}
		\mathbf{K}_2=
		\begin{pmatrix}
			18 (\xi +2 \sigma )^2 \frac{{H}^2 \bar{B}_0^2}{D}  & 0 \\
			0 & \gamma  \frac{k^2}{a^2} \\
		\end{pmatrix}\,,
	\end{equation}
	where
	\begin{align}\label{eq:D-def}
		\begin{split}
			D &\equiv 
			3 \big(\Mpl^2 + 12 \sigma \bar{B}_0^2 \big) H^2
			+ \bar{B}_0^2 \left[2 \bar{B}_0 \left(\bar{B}_0 V''-6 \eta 
			H \right)-V'\right]
			\\&\quad- 2 (\xi +2
			\sigma ) \bar{B}_0^2 {H}^2 \left[  
			\frac{k^2 }{a^2{H}^2} 
			- 6 \bigg(\frac{\dot{{ H}}}{{H}^2}
			- \frac{ \dot{\bar{B}}_0}{{ H} \bar{B}_0} \bigg)
			\right] \,.
		\end{split}
	\end{align}
	Both $\mathbf{K}_{3}$ and $\mathbf{K}_2$ have rank 2 and we need to impose an extra condition to reduce their rank to 1 in order to get rid of an unwanted propagating degree of freedom. Looking at $\mathbf{K}_{3}$, we find the following two possibilities
	\begin{align}\label{eq:DC1}
		&\xi +2\sigma =0  \quad \text{and} \quad \xi +2\sigma+2\varsigma \neq 0 \,,
		\\ \label{eq:DC2}
		&\xi +2\sigma \neq 0 \quad \text{and} \quad \xi +2\sigma+2\varsigma = 0 \,.
	\end{align}
	If we impose \eqref{eq:DC1} in $\mathbf{K}_3$, $\alpha$ becomes non-dynamical and after integrating it out we find that the rank of the resultant $2\times2$ kinetic matrix is 2 unless we impose $\varsigma=0$. Similarly, if we impose \eqref{eq:DC2} in $\mathbf{K}_3$, $\alpha$ becomes non-dynamical again and after integrating it out, we find that the rank of the resultant $2\times2$ kinetic matrix is 2 unless we impose $\xi +2\sigma=0$. Looking at $\mathbf{K}_{2}$, we find
	\begin{align}\label{eq:DC3}
		&\boxed{\xi +2\sigma =0  \quad \text{and} \quad \varsigma = 0 } \,.
	\end{align}
	Therefore, the above condition is the {\it only} choice, which together with \eqref{eq:upsilon-zero}, lead to one propagating scalar mode.\footnote{Issues in the bumblebee model for $\sigma=0$ and $\xi\neq0$ was noticed at the level of background equations in \cite{Bailey:2013fwa}.} In other words, imposing the above condition from the beginning in $\mathcal{L}^{(2)}_2$, we can show that not only $\alpha$ but also $\delta{B}_0$ becomes non-dynamical and we will have one scalar propagating degree of freedom as desired. 
	We therefore conclude that we should impose Eqs~\eqref{eq:upsilon-zero} and \eqref{eq:DC3} to make sure that there is only one propagating degree of freedom in the theory. We summarize the possible couplings and their resulting number of degrees of freedom in Table~\ref{tab:dofs}. This opens a new avenue for constructing healthy higher-order theories, which is the underlying logic behind the generalized Proca theories \cite{Heisenberg:2014rta}, but has been largely overlooked\footnote{Degrees of freedom in the bumblebee model on a flat background were counted in \cite{Seifert:2018mmr,Bailey:2025oun} and stability was discussed in specific subsets in \cite{Bluhm:2008yt}.} in studies of the bumblebee model; for example, see \cite{Ovgun:2018xys,Casana:2017jkc,Li:2020dln,Maluf:2020kgf,Bertolami:2005bh,Ovgun:2018ran,Ding:2019mal,Kanzi:2019gtu,Xu:2025bvx,Li:2025bzo,Khodadi:2025wuw,Liang:2022hxd,Khodadi:2021owg,Gullu:2020qzu,Ding:2020kfr,Reyes:2024hqi,Kostelecky:2003fs}. 
	\begin{table}[h!]
		\centering
		\begin{tabular}{c c c c}
			\toprule
			Operator & Der. order ($D$) & d.o.f. & Correct d.o.f.\\\midrule
			$\eta B^2 \nabla_{\mu}B^{\mu}$ & 1 & 1S+2V+2T & \boxed{\rm YES} \\
			$\varsigma (\nabla_{\mu}B^{\mu} )^{2} $ & 2 & 2S+2V+2T & NO \\
			$ \xi B^\mu B^\nu R_{\mu\nu} $ & 2 & 2S+2V+2T & NO \\
			$\sigma B^2 R $ & 2 & 2S+2V+2T & NO \\
			$\upsilon \nabla_\mu B^\mu R$ & 3 & 3S+2V+2T & NO \\
			$\# B^\mu B^\nu G_{\mu\nu}$ & 2 & 1S+2V+2T & \boxed{\rm YES} \\\bottomrule
		\end{tabular}
		\caption{Couplings and the corresponding number of propagating degrees of freedom (scalar (S), vector (V), and tensor (T)). The rightmost column indicates whether the number of propagating degrees of freedom is correct.}
		\label{tab:dofs}
	\end{table}
	
	Imposing now Eqs~\eqref{eq:upsilon-zero} and \eqref{eq:DC3} in the action \eqref{eq:s2scalar-0} and then integrating out the non-dynamical fields $\{\beta,\alpha,\delta{B}_0\}$, we find the final quadratic action for scalar perturbations as
	\begin{align}
		S^{(2)}_{B,S} = \frac{1}{2}\int \frac{\D^3k}{(2\pi)^3}  \, \D{t} \,a^3 \bigg(\mathcal{K} \delta\dot{B}_s^2- {\cal G} \frac{k^2}{a^2}\delta B_s^2\bigg),
	\end{align}
	where
	\begin{align}\label{eq:K}
		{\cal K}&\equiv 2 \Mpl^2 H^{2}{\cal K}_T \frac{3 \eta  \bar{B}_{0} H \big( \Mpl^2 + 3 \xi  \bar{B}_{0}^2\big)
			+3 \eta ^2 \bar{B}_{0}^4
			+ 2 \bar{B}_{0}^2 \big(6 \xi ^2 H ^2 - \big( \Mpl^2-\xi \bar{B}_{0}^2\big) V''\big)}{\big[\eta  \bar{B}_{0}^3 - H \big(\Mpl^2 - 3 \xi \bar{B}_{0}^2\big)\big]^2} \,,
	\end{align}
	which is presented for the modes deep inside the horizon $k\gg{aH}$.
	The gradient coefficient ${\cal G}\neq 0$ has an involved form; however, as we will see, its explicit expression is not needed here.
	In the expression of ${\cal K}$, we can substitute for $V''$, taking the first derivative of the \eqref{eq:VFE} (imposing the conditions \eqref{eq:upsilon-zero} and \eqref{eq:DC3}), and then substituting for $\dot{\bar{B}}_{0}$ (or equivalently $\dot{H}$) from a linear combination of the first Friedmann equation \eqref{eq:FE1} and the second Friedmann equation \eqref{eq:FE2}, we find 
	\begin{align}\label{eq:K-zero}
		\boxed{{\cal K} = 0} \,.
	\end{align}
	We would like to mention that, although we have presented ${\cal K}$ for the modes deep inside the horizon in Eq.~\eqref{eq:K}, the result Eq.~\eqref{eq:K-zero} holds for the full expression of ${\cal K}$ as well.
	
	Let us elaborate further on the result \eqref{eq:K-zero}. Since the bumblebee model propagates three degrees of freedom in Minkowski spacetime (two transverse vector modes and one longitudinal scalar mode), the condition \eqref{eq:K-zero} cannot be interpreted as an additional degeneracy condition, unlike Eqs.~\eqref{eq:upsilon-zero} and \eqref{eq:DC3}. Indeed, one could show that once one goes beyond linear perturbations a non-vanishing kinetic term could be generated. The fact that the kinetic term vanishes at linear order but reappears at higher order indicates that the scalar mode $\delta B_s$ cannot be treated perturbatively around the linearized theory: we will discuss this in more detail in the next section.
	
	\subsection{Final action}\label{sec:finalaction}
	Based on the linear perturbation analysis above, the results \eqref{eq:upsilon-zero} and \eqref{eq:DC3} are necessary conditions to have correct degree of freedom in cosmological setup. The action \eqref{eq:lagr} then simplifies to
	\begin{equation}\label{eq:lagr-final}
		\begin{split}
			S_B = \int \D^4x \sqrt{-g} 
			\left[ 
			\frac{M_{\rm Pl}^{2}}{2} R - \frac{1}{4}B_{\mu \nu}B^{\mu \nu} - V(B^2) + \eta B^2 \nabla_{\mu}B^{\mu} 
			+ \xi B^\mu B^\nu G_{\mu\nu} \right]
			\,,
		\end{split}
	\end{equation}
	where we have replaced $\sigma$ in favour of $\xi$ using \eqref{eq:DC3}. In the above action we have set $\gamma=1$ without loss of generality.
	The action \eqref{eq:lagr-final} is now a subset of the theory \eqref{eq:lagr}, which itself is a subset of generalized Proca theory~\cite{Heisenberg:2014rta}. Expanding the Einstein tensor in terms of Ricci tensor and using the identity \eqref{eq:identity1}, we find the following dictionary between our model and generalized Proca theories
	\begin{equation}
		\label{eq:PG_map}
		G_2 = -\frac{1}{4}B_{\mu\nu}B^{\mu\nu}-V_B\left(-2X\right), \quad G_3=-2\eta {X}, \quad G_4 = \frac{M_{\rm Pl}^2}{2} + \xi X \,;
		\qquad
		X \equiv -\frac{1}{2} B_\mu B^\mu \,,
	\end{equation}
	with $c_2=0$. In the above dictionary, we have used the notation from Eq.~(2.2) in Ref.~\cite{Heisenberg:2014rta}.
	
	The cosmological background equation for the above model can be found by imposing conditions \eqref{eq:upsilon-zero} and \eqref{eq:DC3} on Eqs~\eqref{eq:FE1}, \eqref{eq:FE2}, and \eqref{eq:VFE} which gives the following compact equation
	\begin{align}
		\big(\Mpl^2 - \xi \bar{B}_{0}^{2} \big) \dot{H} =
		\big(\xi H + \frac{\eta}{2}\bar{B}_0\big)(\bar{B}_0^2)^{\dot{}} \,.
	\end{align}
	The above equation has an exact de Sitter solution when $\xi=\eta=0$, and we note that the bumblebee field acts as a cosmological constant in this limit (this was also noticed in e.g. \cite{Xu:2025bvx} and discussed in \cite{Gonzalez-Espinoza:2025fmi}). Integrating the above equation we find
	\begin{align}
		{H} = \frac{3\Mpl^2{H}_{\rm dS}+\eta\bar{B}_{0}^3}{3 \big(\Mpl^2 - \xi \bar{B}_{0}^2 \big)} \,,
	\end{align}
	where the constant of integration is fixed such that $H=H_{\rm dS}$ for $\eta=0=\xi$. Using the above result, we find
	\begin{align}
		V(-\bar{B}_0^2) = \frac{\big(3 \Mpl^2 {H}_{\rm dS} + \eta \bar{B}_{0}^{3}\big)^{2}}{3 \big(\Mpl^2 - \xi \bar{B}_{0}^2 \big)} \,.
	\end{align}
	This result is highly restrictive, indicating that, once the degeneracy conditions \eqref{eq:upsilon-zero} and \eqref{eq:DC3} are imposed, the potential at the background level is completely fixed by the background equations and is no longer arbitrary in the theory. Since the potential cannot be chosen to be of the form $V=V(B^2-b^2)$, we cannot interpret the model as being spontaneously Lorentz-symmetry breaking. The constant $\Lambda_B\equiv 3 \Mpl^2 H_{\rm dS}^2$ sets the bumblebee scale, and the couplings $\eta$ and $\xi$ quantify the deviation from an exact de Sitter solution with cosmological constant $\Lambda_B$.
	
	For the tensor and vector modes, after imposing the degeneracy condition \eqref{eq:upsilon-zero} and \eqref{eq:DC3}, the requirements for the absence of ghost and gradient instabilities are
	\begin{equation}\label{eq:ghost-gradient-tv}
		\mathcal{K}_T = 1-\xi \tilde{B}_0^2 >0 \,, 
		\qquad
		c_T^2 
		= \frac{1+\xi\tilde{B}_0^2}{1-\xi\tilde{B}_0^2} >0 \,,
		\qquad
		c_V^2 
		= 1 + \frac{2\xi^2\tilde{B}_0^2}{1-\xi\tilde{B}_0^2} > 0 \,.
	\end{equation}
	The above conditions together imply
	\begin{align}
		\label{xi-stability}
		\begin{split}
			0 < \xi\tilde{B}_0^2 < 1 \,;& \qquad \mbox{for}\,\, \xi>0 \,,
			\\
			0 < |\xi|\tilde{B}_0^2 < 1 \,;& \qquad \mbox{for}\,\, \xi<0 \,.
		\end{split}
	\end{align}
	The first case $\xi>0$ leads to the superluminal propagation of both tensor and vector modes $c_T^2>1$ and $c_V^2>1$ while, for $\xi<0$, we find subluminal propagation of the tensor modes $c_T^2<1$ and superluminal propagation of vector modes $c_V^2>1$. In an expanding FLRW background, vector modes decay in the absence of sustained sources, so they become negligible at late times.

	The no-ghost condition and the speed of the propagation given in Eq.~\eqref{eq:ghost-gradient-tv} and using the mapping \eqref{eq:PG_map} can be written as
	\begin{align}\label{eq:K-G}
		\mathcal{Q}_{T} & \equiv \Mpl^2 {\cal K}_{T}= 2 \left( G_{4} -2 X G_{4,X} \right) \,, 
		\qquad c_{T}^{2} = \frac{2G_{4}}{\mathcal{Q}_{T}} \,, 
		\qquad c_{V}^{2} = 1 + \frac{4X G_{4,X}^2}{\mathcal{Q}_{T}} \,,
		\\ \nonumber
		{\cal K} &= 
		{\mathcal{Q}}_T \frac{H^2}{X}\cdot
		\frac{ {\mathcal{Q}}_T \big[3 \bar{B}_{0} H X G_{3,X} - 6 H ^2 \big(G_4+2 X G_{4,X} \big)+2 X^2 G_{2,XX}\big] +3 \big(2 H  G_4- \bar{B}_{0} X G_{3,X}\big)^2 }{\big[\bar{B}_{0} X G_{3,X}+2 H \big(G_4-4 X G_{4,X} \big)\big]^2} \,.
	\end{align}
	The above expressions match those found in~\cite{DeFelice:2016yws}. Once we impose \eqref{eq:PG_map} in Eq.~\eqref{eq:K-G}, it reduces to \eqref{eq:K}, which vanishes upon imposing the background equations as can be seen in Appendix~\ref{sec:app_calKzero}. This result is more general and not restricted to our marginal action, which is a subset of generalized Proca with the identification \eqref{eq:PG_map}. 
	
	\section{No-go theorem}\label{sec:nogo}
	After imposing the necessary conditions to have a healthy theory, i.e. removing extra unwanted degrees of freedom, the action in Eq.~\eqref{eq:lagr-final} is precisely a marginal (constant-coupling) subset of generalized Proca theory \cite{Heisenberg:2014rta}, and the relations \eqref{eq:upsilon-zero} and \eqref{eq:DC3} coincide with the so-called degeneracy conditions. In generalized Proca, these degeneracy conditions guarantee that no extra degree of freedom appears on any background. By contrast, bumblebee models are not designed to satisfy such conditions a priori. The key point of our analysis is that we did not impose degeneracy from the outset: instead, we first constructed the most general bumblebee action built from all diffeomorphism-invariant marginal operators compatible with the bumblebee setup \eqref{eq:lagr} (a systematic completion that has been largely overlooked in the bumblebee literature), and then performed the full linear cosmological perturbation analysis of that general action. This analysis shows that even though bumblebee theories need not be degenerate in general, cosmology forces our hand: without Eqs.~\eqref{eq:upsilon-zero} and \eqref{eq:DC3}, the scalar sector contains an unwanted extra longitudinal mode, so one must impose these degeneracy conditions in order to recover the expected single scalar degree of freedom on an FLRW background. However, once this is done, the remaining scalar becomes pathological, its kinetic coefficient vanishes at the linear level of perturbations, as in Eq.~\eqref{eq:K-zero}, so the perturbative description breaks down. In other words, the theory is driven into the degenerate (generalized-Proca-like) locus in cosmology, and that locus makes the surviving scalar ill-defined.
	
	With the above results and discussion in mind, we can list all the necessary and sufficient conditions for a healthy theory and evaluate them in order of importance as
	\begin{equation*}
		\begin{rcases}
			\text{%
				\parbox{0.85\linewidth}{%
					\begin{enumerate}
						\item[(i)] The most general marginal action $\rightarrow$ Eq.~\eqref{eq:lagr} -- $\checkmark$
						\item[(ii)] Isotropic and homogeneous background $\rightarrow$ $B_\mu=B_0(t)\delta^0_\mu$ $\rightarrow$ Eq.~\eqref{eq:bbpert} -- $\checkmark$
						\item[(iii)] No extra {\it propagating} d.o.f. on FLRW (1S+2V+2T) $\rightarrow$ Eqs.~\eqref{eq:upsilon-zero} \& \eqref{eq:DC3} \\(and Table~\ref{tab:dofs}) -- $\checkmark$
						\item[(iv)] Healthy cosmological perturbations (no ghost or gradient instabilities + no strong coupling) $\rightarrow$ Eq.~\eqref{eq:K-zero} $\rightarrow$ $\mathcal{K}=0$ -- FAILS,
					\end{enumerate}
			}}
		\end{rcases}
		\text{No-go result}
	\end{equation*}
	where we see that since Condition (iv) fails, the action~\eqref{eq:lagr-final} cannot represent a healthy theory on a dynamical cosmological background. 
	
	We can now ask whether it is possible to add new terms to the action in order to cure these pathologies: in generalized Proca (and related degenerate scalar-tensor theories), a common way to address strong coupling problems is the scordatura mechanism, i.e. a controlled, small detuning away from exact degeneracy so that the would-be extra mode only becomes dynamical above a cutoff, while the low-energy perturbations become healthy \cite{Motohashi:2019ymr,Gorji:2020bfl,Gorji:2021isn,DeFelice:2022xvq,Aoki:2021wew}. The crucial difference in the bumblebee setup studied here is that we have intentionally restricted the action to marginal operators with dimensionless couplings; consequently, aside from $\Mpl$ and Hubble parameter $H(t)$, there is no independent high-energy scale that could parametrically separate ``the strong-coupling scale'' from ``the scale at which the extra mode propagates''. As a result, a detuning away from Eqs.~\eqref{eq:upsilon-zero} and \eqref{eq:DC3} is not naturally a soft deformation in the EFT sense: it generically reintroduces the extra mode within the same regime where one is trying to trust cosmological perturbation theory, rather than pushing it safely above a controllable cutoff. Therefore, within the general bumblebee model \eqref{eq:lagr} constructed from marginal operators, Conditions (iii) and (iv) can never be simultaneously satisfied, which is precisely why our result is genuinely a no-go theorem for cosmological perturbations in this general bumblebee framework. We highlight that the degeneracy conditions in Eqs.~\eqref{eq:upsilon-zero} and \eqref{eq:DC3} guarantee that the action is ghost-free on any background, but that the converse statement is not true. Even without imposing any degeneracy conditions, one may find that the theory is healthy around some particular background; however, without imposing Eqs.~\eqref{eq:upsilon-zero} and \eqref{eq:DC3} there is in general no way to guarantee a healthy theory without performing linear perturbations or Hamiltonian analysis. This analysis is yet to be done on several bumblebee models in the literature.

	\section{Summary \& Conclusions}\label{sec:disc}
	In the last decade, a plethora of studies on the bumblebee model has been published. This model, which can be regarded as a vector subset of the SME framework, spontaneously breaks local Lorentz and diffeomorphism symmetries while retaining full observer covariance. Thanks to its relative simplicity, it has become a popular testbed for spacetime-symmetry breaking, yet several important aspects have so far been largely overlooked in the literature. In this paper, we have elucidated some important features of this class of models, specifically concerning the non-minimal coupling to gravity by analyzing their cosmological perturbations. In particular, we extended the standard bumblebee action to include all possible marginal operators together with a general vector potential. Since all operators are marginal, no scale enters the action beyond $\Mpl$ and those encoded in the potential. After studying the background equations for homogeneous and isotropic backgrounds, we extended our study to the linear stability analysis. Since this is a $U(1)$ breaking vector theory non-minimally coupled to gravity, we expect that the model should be propagating two tensor modes, two vector modes, and one scalar mode. We found that the tensor modes and the vector modes are stable.
	
	For the scalar sector, our analysis yielded a clear and robust no-go result. In the most general bumblebee theory constructed from all marginal operators, the scalar perturbations around a spatially flat FLRW background do not generically reduce to a single propagating scalar: depending on the non-minimal and derivative couplings, the scalar sector propagates additional degrees of freedom (the longitudinal mode becomes dynamical, and for one particular marginal operator the system even develops higher-derivative dynamics that increases the number of scalar propagating modes). We traced these extra modes to specific couplings and showed that recovering the expected cosmological field content forces the theory to satisfy some degeneracy conditions between the couplings. However, precisely when these relations are imposed and the unwanted scalar(s) are removed, the potential becomes uniquely fixed by the background equations and the remaining scalar mode becomes ill-defined in the sense that its kinetic term vanishes, so the scalar sector is infinitely strongly coupled already at the linear level of perturbations. Since our setup contains only marginal operators with dimensionless couplings, there is no additional high scale available other than $M_{\rm Pl}$ and Hubble parameter $H(t)$ to separate this strong coupling problem from the regime where cosmological perturbation theory is meant to apply. Consequently, within the general marginal bumblebee framework studied here, one cannot simultaneously obtain the correct number of propagating scalar degrees of freedom and a healthy cosmological scalar perturbation sector. This is the essence of our no-go theorem.
	
	\subsubsection*{Acknowledgements}
	CvdB is supported in part by the Lancaster–Sheffield Consortium for Fundamental Physics under STFC grant: ST/X000621/1. MAG, N.A.N, and MY were financed by the Institute for Basic Science under the project code IBS-R018-D3. M.Y. is supported in part by JSPS Grant-in-Aid for Scientific Research Number JP23K20843. N.A.N is grateful for discussions with Alan Kosteleck\'y and Quentin G. Bailey, as well as for the hospitality of the Indiana University Center for Spacetime Symmetries (IUCSS), where some of these results were discussed. M.C.P is supported by the National Research Foundation of Korea (NRF) through grants RS-2023-NR077094 and RS-2020-NR049598 (Center for Quantum Spacetime: CQUeST).
	
	\appendix	
	
	\section{Covariant equations of motion}\label{app:chis}
	
	The quantities $(\chi_i)_{\mu\nu}$ in Eq.~\eqref{eq:generalized_einstein} read
	\begin{equation}
		\begin{aligned}
			(\mathcal{X}_1)_{\mu\nu} &= -2 B^{\alpha} B^{\beta} R_{\alpha \beta} g_{\mu \nu} + 4 B^{\alpha} B_{(\mu} R_{\nu) \alpha} + 2 B_{(\mu|} \big(\nabla_{\alpha}\nabla^{\alpha} B_{|\nu)} \big) - 2 B_{(\mu|} (\nabla_{\alpha} \nabla_{|\nu)} B^{\alpha}) \\& -2 B^{\alpha}\nabla_{\alpha} \nabla_{(\mu} B_{\mu)} + 2 \nabla_{\alpha}B_{\mu} \nabla^{\alpha}B_{\nu} - 2 \nabla_{\alpha} B^{\alpha} \nabla_{(\mu}B_{\nu)} - 2 \nabla^{\alpha} B_{(\mu} \nabla_{\nu)}B_{\alpha}  \\& +g_{\mu \nu} \big(\nabla_{\alpha} B^{\alpha}\big)^2 + 2 g_{\mu \nu}  B^{\alpha} \nabla_{\beta} \nabla_{\alpha} B^{\beta} + g_{\mu \nu} \nabla_{\alpha}B_{\beta} \nabla^{\beta} B^{\alpha} \,, 
			\\
			(\mathcal{X}_2)_{\mu\nu} &= 2 B_{\alpha } B^{\alpha } R_{\mu \nu } + 2 B_{\mu } B_{\nu } R - g_{\mu \nu } \big( B_{\alpha } B^{\alpha }  R - 4 B^{\alpha } \nabla_{\beta }\nabla^{\beta}B_{\alpha } - 4 \nabla_{\beta }B_{\alpha } \nabla^{\beta }B^{\alpha } ) \\&- 4 \nabla_{\mu }B^{\alpha } \nabla_{\nu }B_{\alpha }  - 4 B^{\alpha } \nabla_{\nu }\nabla_{\mu }B_{\alpha } \,,
			\\
			(\mathcal{X}_3)_{\mu\nu} &= -2 B^{\alpha } B^{\beta } g_{\mu \nu } R_{\alpha \beta } + 4 B^{\alpha } B_{(\mu } R_{\nu ) \alpha } - 4 B_{(\mu |} \nabla_{\alpha }\nabla_{| \nu) }B^{\alpha }  + g_{\mu \nu } (\nabla_{\alpha }B^{\alpha })^{2}  \\&+ 2g_{\mu \nu } B^{\alpha }\nabla_{\beta }\nabla_{\alpha }B^{\beta } \,, \\
			(\mathcal{X}_4)_{\mu\nu} &= 2 B_{\mu } B_{\nu } \nabla_{\alpha }B^{\alpha } + 2 B^{\alpha } B^{\beta } g_{\mu \nu } \nabla_{\beta }B_{\alpha } - 4 B^{\alpha } B_{(\mu } \nabla_{\nu) }B_{\alpha }  \,, \\
			(\mathcal{X}_5)_{\mu\nu} &= 2 R_{\mu \nu } \nabla_{\alpha }B^{\alpha }  - 2 \nabla_{\alpha }\nabla_{\nu }\nabla_{\mu }B^{\alpha } + g_{\mu \nu } \big(  B^{\alpha } \nabla_{\alpha }R + 2 \nabla_{\beta }\nabla^{\beta }\nabla_{\alpha }B^{\alpha } \big)- 2 R_{\mu \beta \nu \alpha } \nabla^{\beta }B^{\alpha } \\&+ 4 R_{(\mu | \alpha } \nabla_{ | \nu ) }B^{\alpha } -  2 B_{(\mu } \nabla_{\nu ) }R + 2 B^{\alpha } \nabla_{\nu }R_{\mu \alpha } \,,
		\end{aligned}
	\end{equation}
	where $A_{(\mu \nu)} \equiv \tfrac{1}{2}(A_{\mu \nu} + A_{\nu \mu})$.
	
	\section{Vanishing of the kinetic coefficient in the scalar sector}\label{sec:app_calKzero}
	Here we show that, the subset of generalized Proca theory that is relevant to the bumblebee model we have considered is also strongly coupled for the longitudinal mode. 
	
	We now view the theory as a subclass of generalized Proca and restrict to the functional sector relevant for the bumblebee map, namely
	\begin{align}\label{eq:app-condition}
		G_{3,XX} = 0 \,, 
		\qquad 
		G_{4,XX} = G_{4,XXX} = 0 \,,
		\qquad
		G_{5} = 0 \,.
	\end{align}
	
	With this assumption, the background equations of motion on the nontrivial branch $\bar{B}_0\neq 0$ can be written as
	\begin{align}
		6 H^2 G_{4} + 2 X G_{2,X} - 6 \bar{B}_{0}H X G_{3,X} + G_{2} = & 0 \label{eq:FF1GP-app} \,, \\
		4 \dot{H} (G_{4} - \bar{B}_0^2 G_{4, X}) + \bar{B}_{0} \dot{\bar{B}}_{0}(\bar{B}_{0} G_{3, X} - 4 H  G_{4, X}) = & 0 \label{eq:FF2GP-app} \,, \\
		6 H X G_{3, X} - \bar{B}_{0} (G_{2,X} + 6 H^{2} G_{4, X})= & 0 \label{eq:constraintGP-app}\,.
	\end{align}
	Here, in writing Eq.~\eqref{eq:FF2GP-app}, we have already used the vector-field background equation \eqref{eq:constraintGP-app}.
	
	Now we take a time derivative of Eq.~\eqref{eq:constraintGP-app}
	and substituting $G_{2, X}$ with Eq.~\eqref{eq:constraintGP-app}, we get
	\begin{align}
		6 X \dot{H} G_{3, X}  + 3 \bar{B}_{0} H (\dot{\bar{B}}_{0} G_{3, X} - 4 \dot{H} G_{4, X }) - \bar{B}_{0}^2 \dot{\bar{B}}_{0} G_{2, XX} = 0 \,.
	\end{align}
	Notice that we have used the conditions Eq.~\eqref{eq:app-condition} to obtain the above result.
	
	Now, substituting for $G_{2,XX}$, in \eqref{eq:K-G}, we get 
	\begin{align}
		{\cal K} \propto 4 \dot{H}(G_{4}-\bar{B}_{0}^{2} G_{4, X} ) + \bar{B}_{0} \dot{\bar{B}}_{0} (\bar{B}_{0} G_{3, X} - 4 H G_{4, X}) \,,
	\end{align}
	which is exactly Eq.~\eqref{eq:FF2GP-app} and, hence, the kinetic coefficient vanishes.\footnote{The explicit expressions can be found in the ancillary files available through \href{https://github.com/nilsanilsson/bumblebee_nogo/}{this link}.}
	
	\bibliographystyle{JHEPmod}
	\bibliography{refs}
	
\end{document}